\documentclass{moriond}

\def\Journal#1#2#3#4{{#1} {\bf #2}, #3 (#4)}

\def\NPB{{\em Nucl. Phys.} B}
\def\PLB{{\em Phys. Lett.}  B}

\def\PRD{{\em Phys. Rev.} D}

\def\JHEP{\em JHEP}
\def\EPJC{{\em Eur. Phys. J} C}

\def\be{\begin{equation}}
\def\ee{\end{equation}}
\def\bea{\begin{eqnarray}}
\def\eea{\end{eqnarray}}

\newcommand{\Photo}{}

\begin{document}
\vspace*{4cm}
\title{Heavy Flavoured Meson Fragmentation Functions in $e^+e^-$ annihilation up to NNLO + NNLL \footnote{Contribution to the proceedings of the $57^{\mathrm{th}}$ Recontres de Moriond 2023, QCD and High Energy Interactions.}}

\author{ Leonardo 	Bonino \footnote{Speaker at the conference.} $^*$, Matteo Cacciari $^\dag$ $^\ddag$ and Giovanni Stagnitto $^*$}

\address{$^*$ Physik-Institut, Universit\"at Z\"urich, Winterthurerstrasse 190, CH-8057 Zürich, Switzerland \\ $^\dag$ Sorbonne Universit\'e CNRS, Laboratoire de Physique Th\'eorique et Hautes  \'Energies, LPTHE, F-75005 Paris, France \\ $^\ddag$ Universit\'e Paris Cit\'e, F-75006 Paris, France}

\maketitle\abstracts{
In this contribution,  we make use of the QCD perturbative fragmentation function formalism to describe the one-particle inclusive fragmentation of a heavy quark produced in $e^+e^-$ annihilation at $\mathcal{O}(\alpha_S^2)$. We perform the computation analytically in Mellin-space.  We resum soft-gluons effects in initial conditions and coefficient functions and perform evolution up to NNLL accuracy, obtaining the first NNLO + NNLL prediction for charm quark production.  We study the impact of different matching schemes and Landau pole prescriptions in soft-gluon resummation, finding significant differences.  We extract simple non-perturbative fragmentation functions for $B$ and $D^*$ mesons by comparing the perturbative prediction with the data from CLEO, BELLE and LEP experiments.  We find that for charm mesons the experimental results from CLEO/BELLE and from LEP are not reconcilable with the standard DGLAP evolution.}

\section{Introduction}

The perturbative fragmentation function formalism is used in QCD to resum observables describing the fragmentation of heavy quarks into hadrons.  We consider the \textit{one-particle inclusive} production of a single heavy hadron $H$ ($D$ or $B$ meson) at high energy in $ e^+ + e^- \to V(Q) \to H(p) + X$,  inclusive on all other final state particles $X$, and we study it to next-to-next-to-leading order accuracy in QCD. A paper describing these findings is in preparation \cite{bcs}.

\subsection{The Fragmentation Function formalism}
If the quark is heavy, the \textit{parton fragmentation function} can be factored in perturbative $D^p$ and non-perturbative $D^{np}$ components such that the hadron cross section is $\sigma_H(x,Q^2)=\hat{\sigma}\otimes D^{p}\otimes D^{np}$, with $x$ energy fraction of the heavy quark, ratio of kinematic invariants and $\hat{\sigma}$ massless partonic cross section.  The \textit{process-independent} $D^p$ can be written as \textit{initial conditions} $D^{\mathrm{(ini)}}_{b/Q}$ evolved via DGLAP from the initial factorization scale $\mu_{0F}$, to the final one $\mu_{F}$ \cite{mn}, such that the only \textit{process-dependent} ingredients are the massless partonic \textit{coefficient functions} $C_{a}^{(proc)}$.  It is customary to move to Mellin space where the cross section reads
\begin{equation}
\frac{1}{\sigma_{tot}}\sigma_N\left(Q^2,m^2\right)=\frac{\sigma^{(LO)}}{\sigma_{tot}}\sum_{a,b}C_{a,N}^{(e^+e^-)}\left(Q^2,\mu^2,\mu_F^2\right)E_{ab,N}\left(\mu_F^2,\mu_{0F}^2\right)D_{b/\mathcal{Q},N}^{\mathrm{(ini)}}\left(\mu_0^2,\mu_{0F}^2,m^2\right)\,  ,
\end{equation}
with $\mu$ renormalisation scale, $\sigma_{tot}$ total heavy quark cross section and $\sigma^{(LO)}$ massless parton leading order cross section.  The initial conditions are available analytically in Mellin space up to NNLO \cite{mm,m,mruz},  as well as the coefficient functions  \cite{rv,br,mm2}. We perform the evolution at NNLL accuracy with the public library $\mathtt{MELA}$ \cite{bcn}. 

\subsection{Soft-Gluon Resummation}
In the Sudakov region ($x\to 1$), large logarithms of soft origin appear such that initial conditions and coefficient functions must be resummed \cite{cc,ruz}.  Mellin space analytical expressions for the resummation have recently become available up to NNLL \cite{mruz,acf,cgmp}.  In our calculation, we have implemented both additive and multiplicative matching (such as $\log$-$R$) to match to the fixed order result. A singularity in the resummed initial conditions, giving rise to a branch cut and related to the divergent behaviour of the running coupling $\alpha_S(\mu_0^2)$ near the \textit{Landau pole} $N^L_{\mathrm{ini}}$ at $\mu_0\simeq \Lambda_{QCD}$, signals the onset of non-perturbative phenomena at large $N$ \footnote{For the charm quark at $\sqrt{Q^2}=10.6 \, \mathrm{GeV}$ $N^{L}_{\mathrm{ini}}\sim 7$ and for the bottom at $\sqrt{Q^2}=91.2 \, \mathrm{GeV}$ $N^{L}_{\mathrm{ini}}\sim 32$.}.  To have phenomenologically sensible perturbative predictions a prescription for the pole must be used.  The CNO \cite{cno} prescription accounts to a shift in $N$ with a tunable parameter $f$, while the CGMP \cite{cgmp} prescription truncates the exponent of the Sudakov factor \footnote{While the CNO prescription does not introduce power corrections larger than generally expected for the processes in question, this is the case for the CGMP prescription.}. In Section \ref{tp} we study the effects on phenomenology of the prescription chosen. 

\section{Towards phenomenology} \label{tp}
The non-perturbative part of the fragmentation function is inferred from a fit to the data.  NLO + NLL fits for both $B$ and $D$ hadrons are available \cite{cno} while an effective $\alpha_S$ approach \cite{cf} was calling for a full NNLO analysis.  Recently the NNLO + NNLL level was reached for the bottom quark \cite{cgmp} and fits were performed as well.  Here \cite{bcs} we present $B$-hadrons fits with our implementation and the first fits to the $D$-mesons fragmentation functions at NNLO + NNLL level.  Many experiments throughout the years measured one-particle inclusive observable in $e^+e^-$ collisions.  For $B$-hadrons at the $Z^0$-peak we used data from ALEPH \cite{aleph}, SLD \cite{sld}, OPAL \cite{opal} and DELPHI \cite{delphi} \footnote{While the ALEPH set refers specifically to $B$-mesons, the SLD,  DELPHI and OPAL data are for all weakly decaying bottom-flavoured hadrons and include about $10\%$ of baryons. }. For $D$-mesons the data are available at $\sqrt{Q^2}=91.2\, \mathrm{GeV}$ from ALEPH \cite{alephc} and at $\sqrt{Q^2}=10.6\, \mathrm{GeV}$ from BELLE \cite{belle} and CLEO \cite{cleo1,cleo2}.  We fit $B$ data with a two-parameters non-perturbative ansatz and $D$ data with a four-parameters one as in \cite{cn,cno}. 

\subsection{Charm ratio}
\begin{figure}
\begin{center}
\includegraphics[width=0.43\linewidth]{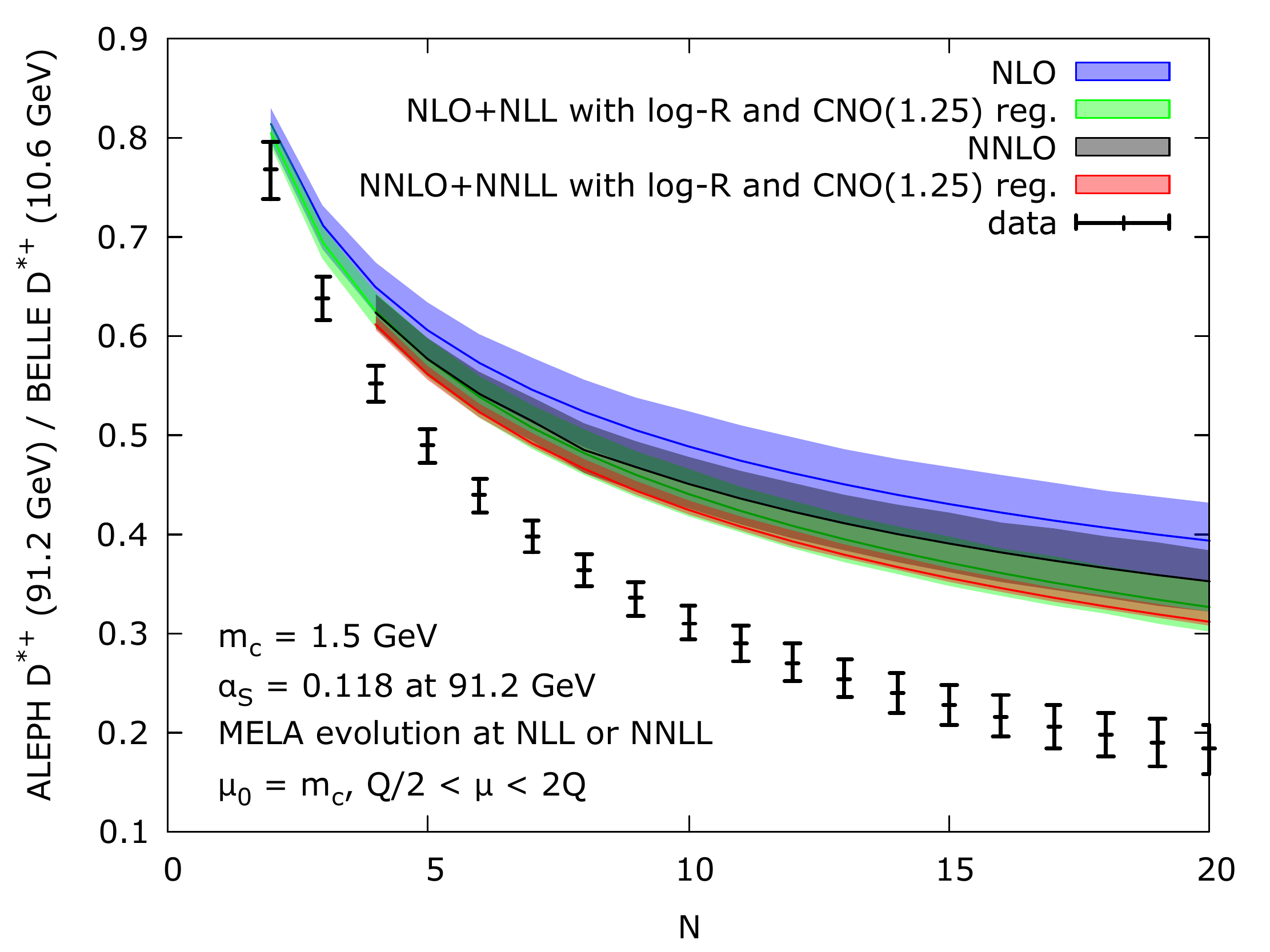}
\end{center}
\caption{Charm ratio up to NNLO + NNLL. Theory band uncertainty: $\mu_{Z/\Upsilon}=\xi  M_{Z/\Upsilon}$ with $1/2<\xi<2$.}
\label{fig:cratio}
\end{figure}

A fully-perturbative observable, the ratio of charm meson moments from LEP and from CLEO and BELLE, was studied at NLO + NLL \cite{cno}, showing an inconsistency between theory and experiments.  In Fig. \ref{fig:cratio} we study this observable at NNLO + NNLL, confirming the inconsistency.  We remark that this observable and the prediction exclusively depend upon the evolution between the $\Upsilon(4S)$ and $Z^0$ energies.  A possible explanation for the inconsistency can be found in the presence of large non-perturbative power suppressed effects in the coefficient functions.  

\subsection{Effects of Landau Pole prescription}
\begin{figure}
\centering
\begin{minipage}{0.49\linewidth}
\includegraphics[width=0.9\linewidth]{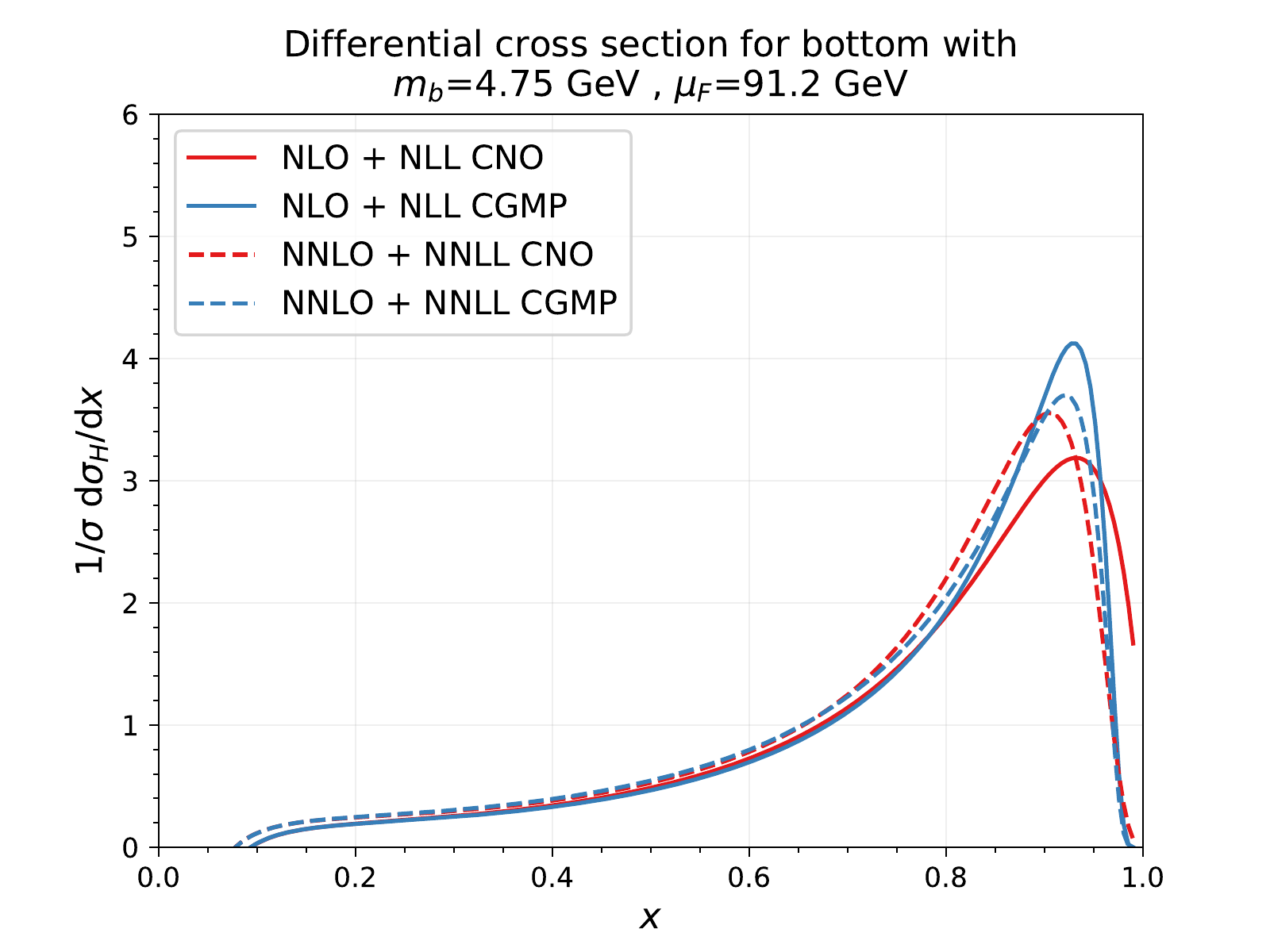}
\end{minipage}
\begin{minipage}{0.49\linewidth}
\includegraphics[width=0.9\linewidth]{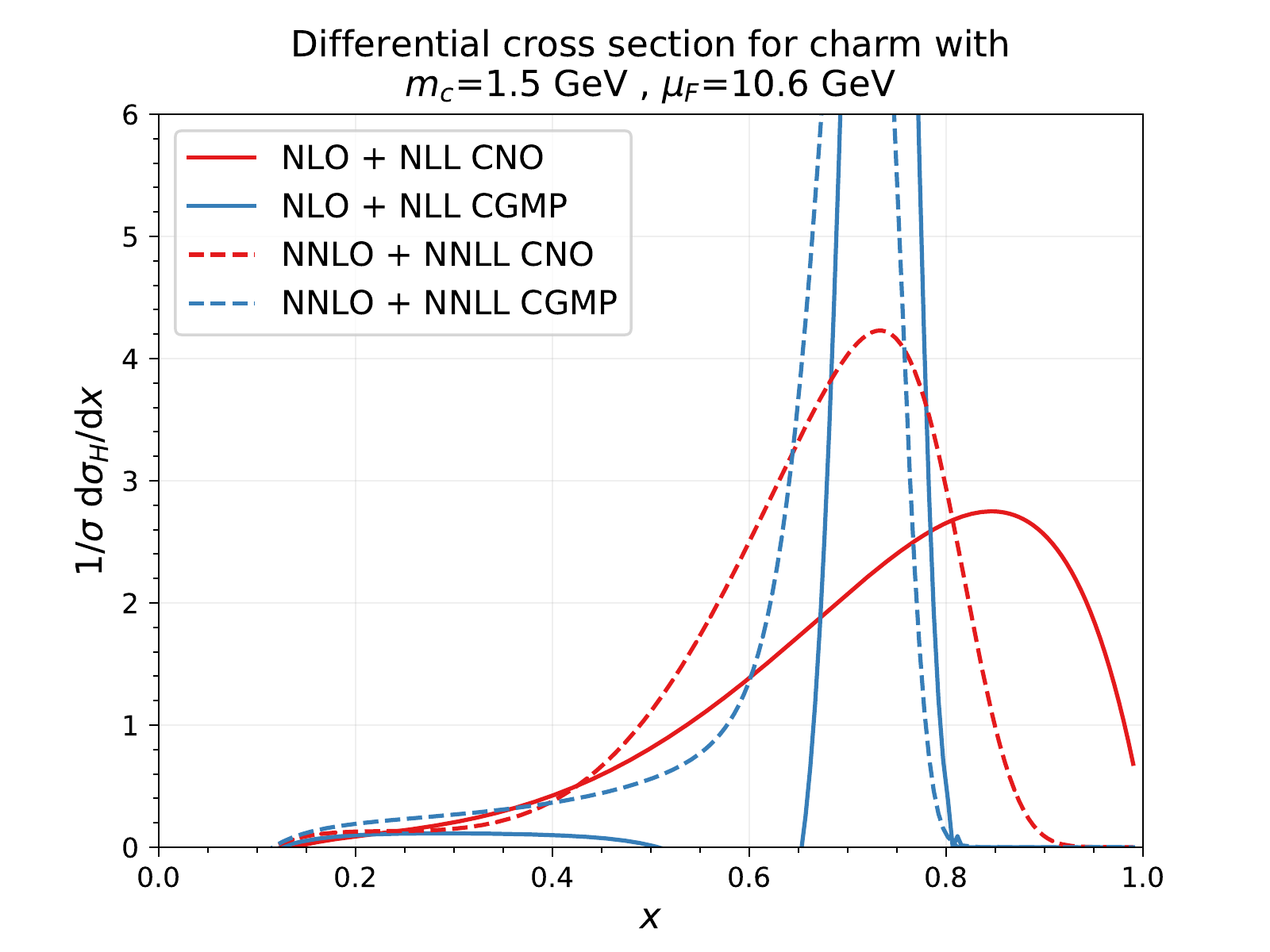}
\end{minipage}
\caption{Left: Bottom differential cross section al NLO + NLL and NNLO + NNLL with CNO ($f=1.25$) and CGMP prescriptions for $\mu_F=\sqrt{Q^2}=91.2\, \mathrm{GeV}$ in $\log$-$R$ matching scheme.  Right: Same for the charm for $\mu_F=\sqrt{Q^2}=10.6\, \mathrm{GeV}$.}
\label{fig:lp}
\end{figure}

We have also studied the effects of changing the Landau pole prescriptions on the initial conditions \footnote{On the coefficient function the effects of the prescription are negligible since the pole appears at larger values of $N$.}.  We observe a variation much larger than any scale variation and, regardless of the prescription chosen, the result is also unstable under variation of the matching scheme. We conclude that on the initial conditions, the characteristics of the perturbative result are strongly influenced by the choice of the Landau pole regularization. The impact of this effect on the final cross section is shown in Fig. \@\ref{fig:lp} left for the bottom.  Due to the smaller mass (and low experiment scale of $\sqrt{Q^2}=10.6\, \mathrm{GeV}$), the effect is much more severe for the charm as in Fig. \@\ref{fig:lp} right.  We notice that the CGMP prescription is not suited to describe the $D$-meson data already at NLO level with our non-perturbative ansatz, due to the steep fall at $x\sim 0.8$.  Since many data points are in that region, for all our fits the CNO prescription was used. 

\subsection{Fits to the data}
As expected, the impact of the effect seen in Fig. \@\ref{fig:lp} on phenomenology is significant and absorbed in a different results of the fit.  
\begin{figure}
\centering
\begin{minipage}{0.47\linewidth}
\includegraphics[width=0.85\linewidth]{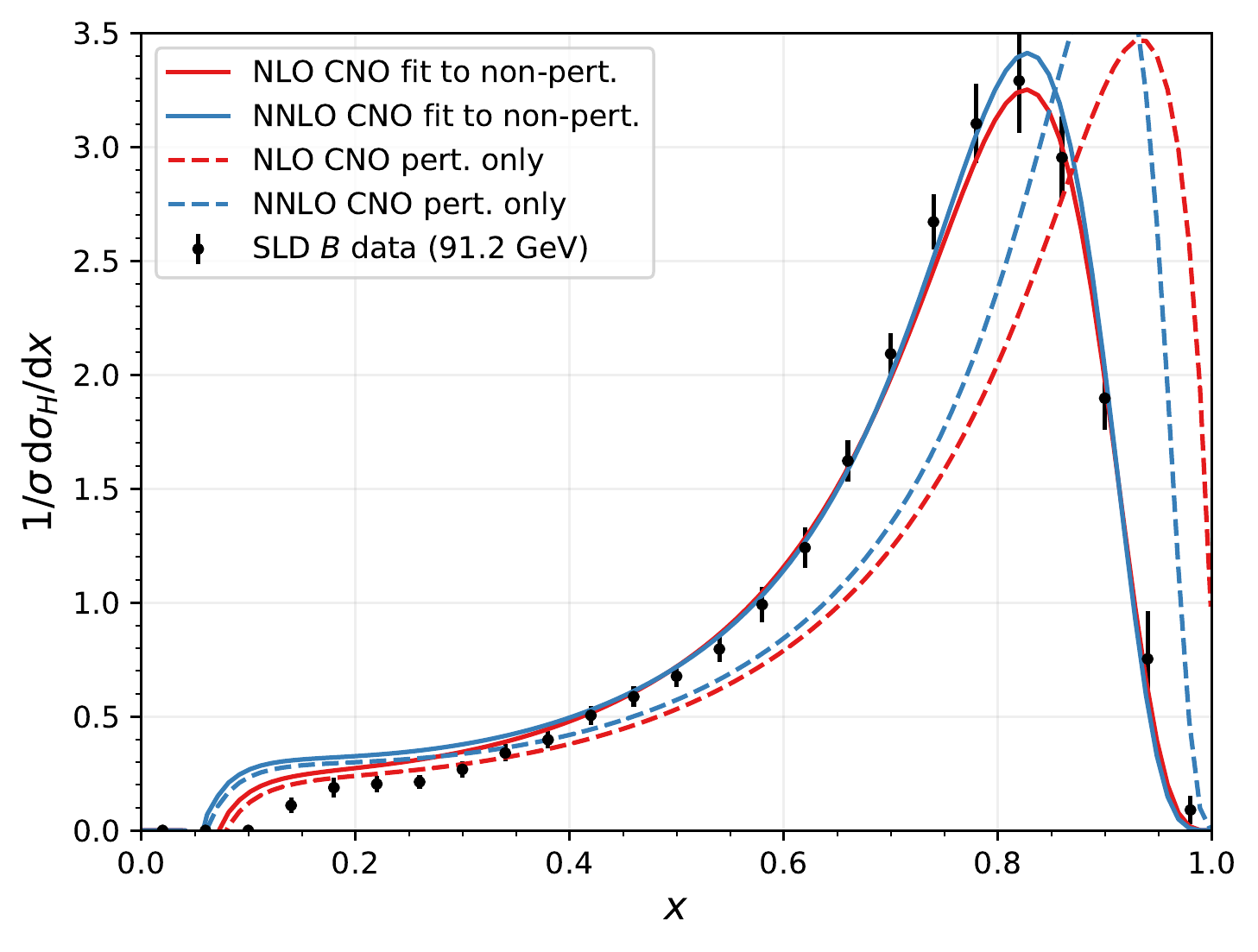}
\end{minipage}
\begin{minipage}{0.47\linewidth}
\includegraphics[width=0.85\linewidth]{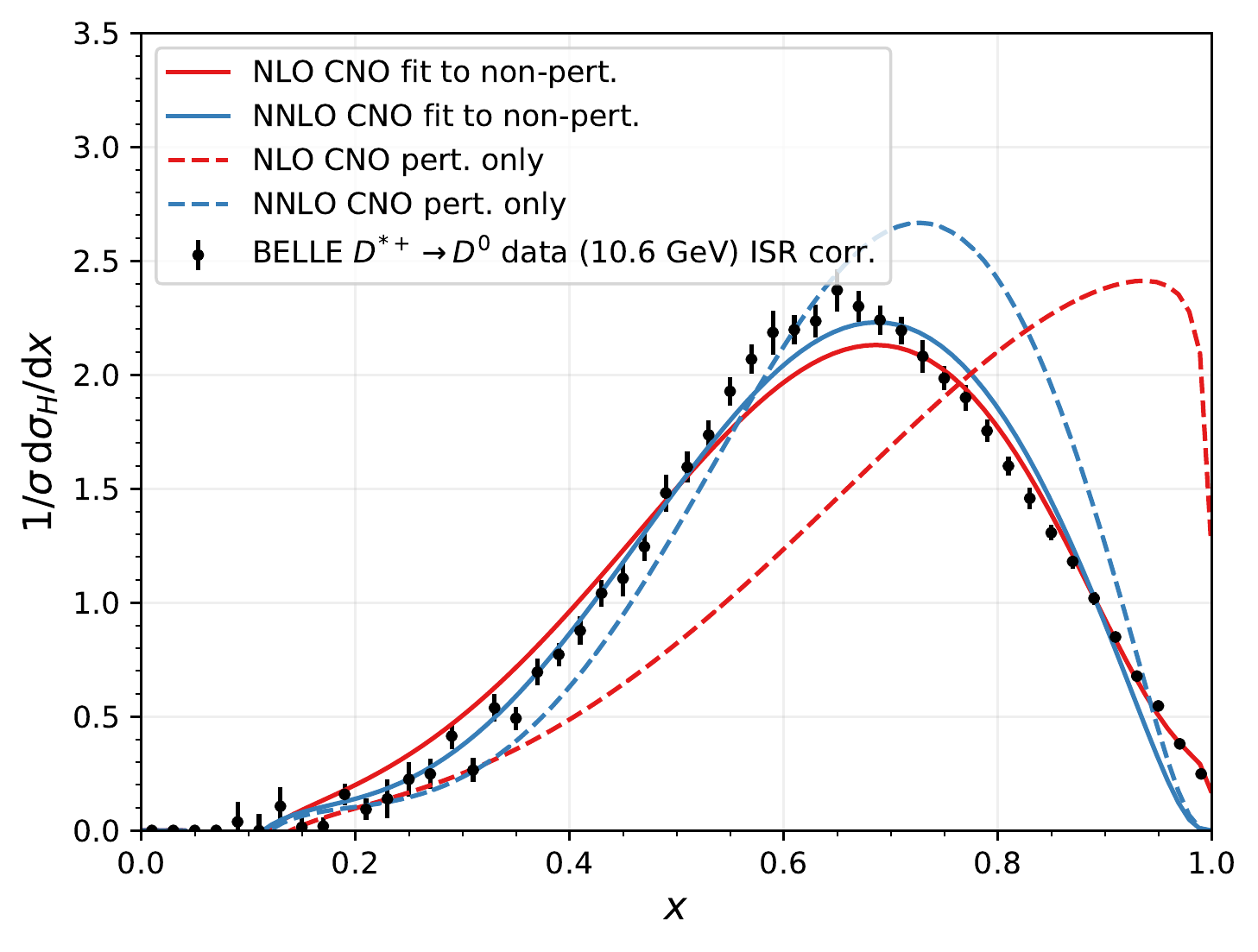}
\end{minipage}
\caption{Left: Fits to SLD data with CNO ($f=1.25$) prescription and $\log$-$R$ matching.  Right: Same for BELLE data ($f=2.0$).}
\label{fig:fits}
\end{figure}
In Fig. \@\ref{fig:fits} our results of the fits to the $B$ and $D$ hadrons fragmentation functions are presented for two of the data sets.  While all fits for the $B$-hadrons are satisfactory,  at NNLO + NNLL level we are not able to describe the $D$-mesons data in the $x> 0.9$ region (for all data sets) as shown in Fig. \@\ref{fig:fits}. We suspect the reason for this being related to the Landau pole non-perturbative effects.

\section{Conclusions}
We have presented the first results for the NNLO + NNLL one-particle inclusive charm fragmentation in $e^+e^-$ annihilation and relative fits to the data.  We uncover large non-perturbative effects arising from the choice of the prescription for Landau pole.  The characteristics of the perturbative result are strongly influenced by the choice of the Landau pole regularization, which has a direct impact on the appropriate shape of the non-perturbative component needed to describe experimental data. Furthermore, the inclusion of one more perturbative order confirms the charm ratio inconsistency.  In conclusion,  non-perturbative effects, possibly of different origin, play a crucial role in precision observables for charm quark fragmentation.

 \section*{Acknowledgments}

LB would like to thank Thomas Gehrmann for useful discussions. This work has received funding from the Swiss National Science Foundation (SNF) under contract 200020-204200.

\end{document}